# Transition from non-sequential to sequential double ionisation in many-electron systems

Michael G. Pullen[1,*], Benjamin Wolter[1], Xu Wang[2], Xiao-Min Tong[3], Michele Sclafani[1], Matthias Baudisch[1], Hugo Pires[1], Claus Dieter Schröter[4], Joachim Ullrich[4,5], Thomas Pfeifer[4], Robert Moshammer[4], J. H. Eberly[6] and Jens Biegert[1,7]

1 ICFO - Institut de Ciencies Fotoniques, The Barcelona Institute of Science and Technology, 08860 Castelldefels (Barcelona), Spain

2 J.R. Macdonald Laboratory, Physics Department, Kansas State University, Manhattan, Kansas 66506-2604, USA

3 Center for Computational Sciences, University of Tsukuba, Tsukuba 305-8577, Japan

4 Max-Planck-Institut für Kernphysik, Saupfercheckweg 1, 69117 Heidelberg, Germany

5 Physikalisch-Technische Bundesanstalt, Bundesallee 100, 38116 Braunschweig, Germany

6 Center for Coherence and Quantum Optics, University of Rochester, Rochester, New York 14627, USA

7 ICREA - Institució Catalana de Recerca i Estudis Avançats, 08010 Barcelona, Spain

* michael.pullen@icfo.eu

## Abstract

Obtaining a detailed understanding of strong-field double ionisation of many-electron systems (heavy atoms and molecules) remains a challenging task. By comparing experimental and theoretical results in the mid-IR regime, we have unambiguously identified the transition from non-sequential (e,2e) to sequential double ionisation in Xe and shown that it occurs at an intensity below $10^{14}$ Wcm$^{-2}$. In addition, our data demonstrate that ionisation from the Xe 5s orbital is decisive at low intensities. Moreover, using the acetylene molecule, we propose how sequential double ionisation in the mid-IR can be used to study molecular dynamics and fragmentation on unprecedented few-femtosecond timescales.

## Main text

The interaction of short ($\tau < 100$ fs) and strong laser fields ($10^{13}$ - $10^{16}$ Wcm$^{-2}$) with matter drives a continually growing range of research fields and applications. Attosecond science [1] and laser-induced electron diffraction [2], for example, are based on the generation of high energy and short duration photon and electron pulses, respectively, using intense near-IR and mid-IR radiation as driving fields. The accepted model to describe these processes is the well-known rescattering scenario [3]: ionisation of the electron near the peak of the electric field followed by acceleration and subsequent re-collision with the parent ion roughly three quarters of a cycle later. Even though this model has been extremely successful at explaining a plethora of observations, new and exciting findings continue to be uncovered that help to further develop [4, 5], and in some cases seriously challenge [6], theoretical interpretations.

One aspect of strong-field physics that continues to provide many insights is double ionisation (DI). Early experiments showed that the detected yield of doubly charged ions below the saturation threshold for ionisation was much higher than expected from two sequential tunnelling events [7]. Modern interpretation is that the enhanced yield of the dication is caused by non-sequential double ionisation (NSDI), which is intricately linked to the re-collision of the first electron (e$_1$) and can proceed via two routes: the second electron (e$_2$) can either be directly ionised by e$_1$ via the so-called (e, 2e) mechanism or it can be resonantly excited by e$_1$ and subsequently tunnel ionised at a later time (RESI). Double ionisation is a complex process that is known to be dependent on a number of laser parameters such as the intensity [8], polarisation [9], pulse duration [10] and wavelength [11]. NSDI is observed below the saturation intensity ($I_S$), which is the point where sequential double ionisation (SDI) starts to dominate, and seems to be ubiquitous in low-Z atomic targets [12] and small molecules [13]. Interestingly, in the case of high-Z atoms (Z: nuclear charge) such as Xe, which is one of the most well studied atomic species, the role of NSDI is not as clear [14]. In the near-IR (0.8 μm < λ < 3 μm) wavelength regime different enhancement mechanisms such as multiple electron re-collisions [15, 16] and resonant excitations [14, 15, 17] have been proposed to explain the disparate results. Nowadays it is generally accepted that ion yield measurements alone are not sufficient to unambiguously determine the regime of double ionisation. Additional measurements of the doubly charged ion momentum distribution [8] or electron correlations [8, 18] are required. The most sophisticated investigation of DI in Xe utilised these techniques and showed that, unlike low-Z targets, the results were suggestive of SDI at the low peak intensities where NSDI normally dominates [19]. Screening of the valence electrons by core electrons was proposed to explain the surprising results.

It is often beneficial to scale experiments to longer wavelengths such that tunnelling regime conditions can be approached and the experimental results can be accurately described using classical methods. Two results have been reported where the goal was to investigate DI in Xe at longer wavelengths [20, 21]. Gingras *et al.* explored a wide wavelength range while monitoring single and double ion yields as a function of the laser intensity. In addition to supporting the idea of resonances at shorter wavelengths, there was evidence for the occurrence of NSDI at longer wavelengths. The position of the famous 'knee structure', which is traditionally near the meeting point of SDI and other mechanisms, was observed to be wavelength dependent. Indeed, as the wavelength was increased from 1 μm to 2 μm the appearance intensity decreased by a factor of two to $1 \times 10^{14}$ Wcm$^{-2}$. If these results represent the general trend then they suggest that SDI should become dominant below $10^{14}$ Wcm$^{-2}$ in the mid-IR (λ > 3 μm). Interestingly, this expectation was not observed at wavelengths of 3.2 μm and 3.6 μm where the measured yields for double ionisation could still be described using inelastic electron impact cross-sections [21] and not using Ammosov, Delone, and Krainov (ADK) tunnelling rates [22] as would expected in the SDI regime. Neither of the above reports included doubly-charged ion momentum distributions or electron correlation maps, which as stated above, are generally required to explicitly determine the double ionisation regime. Therefore, no single experiment has unambiguously shown the transition from NSDI to SDI in Xe and there is still much uncertainty and debate as to the mechanisms of double ionisation in high-Z targets.

Here, using ion yields, ion momenta and electron correlations we perform a thorough investigation of the double ionisation dynamics of Xe in the mid-IR regime and unambiguously identify the transition from (e, 2e) NSDI to SDI. By comparing our experimental results with semi-classical calculations, we show that SDI starts to dominate over the (e, 2e) NSDI mechanism well below $10^{14}$ Wcm$^{-2}$, which is at odds with previous reports. These results suggest that important mechanisms that were previously observed in the near-IR are not important in the mid-IR. A comparison with time dependent density functional theory (TDDFT) also shows that the 5s orbital plays an important role during NSDI at intensities close to $10^{13}$ Wcm$^{-2}$, confirming previous observations [16, 17, 20]. Using the molecule acetylene ($C_2H_2$) as an example, we additionally propose how the SDI region can be harnessed in the future to image nuclear dynamics using mid-IR based laser-induced electron diffraction.

We compare our experimental results to semi-classical and time dependent density functional theory (TDDFT) calculations. The semi-classical method assumes tunnelling of the first electron according to ADK rates [22] and the second electron is placed at the vicinity of the ion core with an energy of -20.98 eV (the negative of the second ionisation potential of Xe). This energy includes the kinetic energy, the ion core Coulomb attraction energy and the electron-electron repulsive energy, although the last term is in general small due to the relatively large distance between the two electrons. Given this energy, the position and momentum of the second electron are randomly assigned and the trajectories of both electrons are determined by integrating the time-dependent Newtonian equation of motion [23, 24]. Intensities below 4 x $10^{13}$ Wcm$^{-2}$ can not be investigated with this method due to computational demands. The TDDFT method utilises pseudo-potentials [25] and to compensate spurious self-interaction a positive charge background is added [26], which provides the correct Coulomb tail. An ionisation potential of 12.13 eV is used, which is close to the experimental value. The double ionisation probability is extracted from the ionisation probability of each orbital [27].

To ionise Xe we use a highly stable (1% rms fluctuations over 12 hours) and intense 160 kHz mid-IR source that operates at a wavelength of 3.1 μm [28] and is well suited to explore strong-field physics [29]. Upon focussing using a 50 mm focal length mirror intensities above $10^{14}$ Wcm$^{-2}$ can be achieved. The focussed radiation intersects a cold beam of atomic Xe that has passed two skimming stages into a reaction microscope (ReMi) spectrometer [30], which has a base pressure below $10^{-10}$ mbar without gas load. The laser polarisation direction is parallel to the electric and magnetic fields of the ReMi spectrometer, which means that longitudinal momentum information can be directly inferred from time-of-flight (TOF) spectra.

In Fig. 1a we present the results of monitoring Xe$^+$-Xe$^{4+}$ ion yields as a function of the peak intensity. The single ionisation (circles) results show the typical saturation-like behaviour as the intensity is increased towards $I_S$ ~ 5.0 x $10^{13}$ Wcm$^{-2}$ [31]. The doubly charged ion (squares) data show a similar gradient for intensities below $I_S$, as is generally observed in the NSDI regime [32]. Rescaled single ionisation data indicated by the solid black curve clearly illustrates this similarity. Approaching $I_S$ NSDI yield begins to plateau in a similar way to the single ion yield due to the lack of neutral atoms in the interaction volume [13]. At intensities higher than $I_S$ ,however, the Xe$^{2+}$ yield starts to increase again and this trend continues up until the maximum intensity of 1.2 x $10^{14}$ Wcm$^{-2}$. Such a pronounced change in the intensity dependence of the ion yield is characteristic of the transition from NSDI to SDI [32]. The yields of the triple and quadruple charged ions (triangles and diamonds, respectively) do not show this trend. The increase in Xe$^{2+}$ counts is more obvious when viewed as a ratio of doubly to singly charged ion yield, as presented in Fig. 1b (squares). A plateau at a value between 4-5 x $10^{-3}$ is observed up until $I_S$, upon which the ratio starts to increase drastically. We note that while the Xe$^{3+}$ and Xe$^{4+}$ data reported in Ref. [21] for similar wavelengths agree well with our observations, the trend in doubly-to-singly charged ion yield ration definitively does not. The reason for this isn't clear but it could be related to the intensity calibration, the accuracy of which is generally limited to the tens of percent level or worse [33].

The semi-classical calculations (solid black), which have been rescaled by a factor of 0.3 to compensate for the absence of focal averaging, reproduce both the plateau and the gradient of the Xe$^{2+}$/Xe$^+$ ratio above $I_S$. This comparison clearly shows that double ionisation in the mid-IR regime can be accurately modelled using semi-classical methods. The TDDFT results for intensities between 2-5 x $10^{14}$ Wcm$^{-2}$ are also presented (dashed black) and emphasise the decisive role of the 5s orbital for NSDI in the mid-IR regime: The calculations that include the 5s orbital reproduce the change in ratio observed for lower intensities very well, whereas those that do not account for its influence drastically differ.

The evolution of the TOF spectra of Xe$^{2+}$ ions as a function of intensity is presented in Fig. 1c. Each spectrum contains five isotopes and is normalised before being vertically shifted for visibility. Apart from the lowest intensity, a clear transition from a double-peak structure at low intensities (bottom traces) to a single peak at high intensities (upper traces) is observed for each isotope. The double peak structure is suggestive of the (e, 2e) NSDI mechanism where the doubly charged ion is created at a phase close to the zero-crossing of the electric field yielding a large drift momentum [8]. To the best of our knowledge, a double hump structure in the $Xe^{2+}$ momentum distribution has never been reported in the literature before. The results seem to rule out an influence of the RESI and resonant enhancement mechanisms as both are known to 'fill the valley' in between double hump [8, 19]. The absence of these mechanisms is not unexpected since: 1) RESI dominates when the returning electron energies are large enough to excite the ion but too low for impact ionisation, which is not the case here, and 2) the contribution from resonant enhancement decreases with increasing wavelength [20]. The narrowing of the ion momentum distributions as the intensity is increased indicates a convergence towards purely sequential ionisation [8]. The same analysis of the $Xe^{3+}$ TOF (not presented here) spectra shows that the double hump behaviour persists at the highest intensity, which indicates that non-sequential ionisation is still the dominant mechanism for creating triply-charged ions.

The excellent agreement of the semi-classical yield ratios in Fig. 1b with experiment provides confidence that the calculations capture the interaction mechanisms and thus opens up the ability to track the individual electron trajectories in order to get further insights. In Figs. 2a,b, we present representative trajectories for both ionised electrons for intensities of 4.0 x $10^{13}$ Wcm$^{-2}$ (left column) and 1.2 x $10^{14}$ Wcm$^{-2}$ (right column). For the lower intensity we see that $e_1$ (dashed purple line) is emitted at the peak of the pulse (t = 0 cycles) and is able to directly impact-ionise $e_2$ (solid yellow line) upon returning to the core. This is a typical example of (e, 2e) double ionisation where $e_2$ is ionised at the moment of re-collision and the two electrons show correlated behaviour by leaving in the same direction. In this particular example $e_1$ approaches the core twice but only re-collides on the second pass (see inset). The fact that the experimental results can be reproduced without including excitations or resonances confirms that these mechanisms are not important in this regime. The motions of the two electrons at the higher intensity are tellingly different. Due to the much higher intensity, $e_1$ can be emitted much before the peak intensity and does not at all return to the vicinity of the core. For $e_2$ to also be emitted it must undergo a sequential ionisation process in which case little correlation between the two electrons is expected. Interestingly, in this example $e_2$ returns to the vicinity of the parent dication half a cycle later.

Correlations between $e_1$ and $e_2$ can be deduced by comparing their theoretical (Figs. 2c,d) and experimental (Figs. 2e,f) longitudinal momenta. For the lower intensity both show a pronounced correlation that manifests itself by most counts predominantly occupying the first and third quadrants. A slight fork-like structure is observed along the diagonal for both experiment and theory, indicating the excellent agreement between the two. It is interesting to note that both show counts in the second and fourth quadrants as well, which indicates that there is a small amount of anti-correlated electron emission. These results are distinctly different from a recent experiment where no correlations were found for any of the investigated intensities [19]. In fact, apart from some initial evidence presented by our group [29], the observation of electron correlations during double ionisation of Xe does not seem to exist in the literature. For the higher intensity no evidence of electron correlations are present in either the experiment or simulations, indicating that this intensity is well within the SDI regime.

The above results show that the trajectories of both electrons resulting from double ionisation in the mid-IR regime can be accurately modelled using a simple classical approach [Ho2005]. This means that any elastic rescattering event that occurs during SDI in the mid-IR can be interpreted classically as is the case in laser-induced electron diffraction (LIED), which is a technique that can image molecular structure and dynamics [34]. If the omnipresent fragmentation processes that accompany molecular ionisation can be shown to be associated with sequential ionisation then they can potentially be imaged using the LIED technique.

To investigate this proposal we present acetylene ($C_2H_2$) ion yield ratios as a function of the laser intensity in Fig. 3a. The $C_2H_2^{2+}/C_2H_2^+$ trend (circles) is very similar to the Xe case, suggesting the dominance of SDI at intensities above $I_S \sim 4 \times 10^{13}$ Wcm$^{-2}$. The decrease in $I_S$ relative to Xe is due to the 0.7 eV lower ionisation potential. In order to confirm the SDI mechanism we show in Fig. 3b how the $C_2H_2^{2+}$ longitudinal momentum distribution evolves from a double peak structure into a single peak as the intensity is increased. The inset illustrates the method to determine $I_S$ from the ion yield measurements [31]. The $C_2H_2$ dication has a well-known deprotonation channel ($C_2H_2^{2+}$ -> $H^+ + C_2H^+$) that progresses via the $\Pi_u$ excited state [35] that also shows the onset of sequential behaviour (squares) at intensities near $\sim$3-4 $\times 10^{13}$ Wcm$^{-2}$. We have confirmed that the majority of the H+ counts result from dissociation of the dication, and not the dissociation of the cation, by performing a coincidence measurement at an intensity of 6 $\times 10^{13}$ Wcm$^{-2}$. Therefore, ultrafast fragmentation processes such as deprotonation from $C_2H_2^{2+}$ excited states, which cannot be temporally resolved using other imaging techniques due to their few-femtosecond nature, could be imaged by taking advantage of SDI within the mid-IR wavelength regime.

We have shown for the first time that strong-field ionisation of Xe in the mid-IR progresses from NSDI via the (e, 2e) mechanism at intensities near $10^{13}$ Wcm$^{-2}$ to SDI as the intensity is increased towards $10^{14}$ Wcm$^{-2}$. This result contradicts previous reports and shows that mechanisms that have previously been shown to be important at shorter wavelengths become negligible in the mid-IR. The influence of the 5s orbital is shown to influence the NSDI yield dramatically at intensities close to $10^{13}$ Wcm$^{-2}$. We also propose that SDI in the mid-IR can be utilised to open a new field of research: imaging of few-femtosecond molecular fragmentation dynamics using the LIED technique. This possibility is extremely enticing as it represents the possibility to image molecular fragmentation channels on timescales that are unprecedented when using other imaging methods.

# Figure captions

**Figure 1** (a) The number of $Xe^+$ (circles), $Xe^{2+}$ (squares), $Xe^{3+}$ (triangles) and $Xe^{4+}$ (diamonds) ions detected as a function of the estimated peak laser intensity. The $Xe^+$ data is also scaled to overlap with the $Xe^{2+}$ curve at intermediate intensities (black line). The circled points represent the intensities where the data in Fig. 2 was acquired. The estimated ±20% error in the absolute intensity determination is indicated on the first and last $Xe^+$ data points. (b) The corresponding ion yield ratios shown alongside the semi-classical (solid black) and TDDFT (dashed black) calculations. (c) Normalised and vertically shifted TOF spectra for five Xe double ion isotopes as the intensity is increased from bottom to top.

**Figure 2** (a) & (b) Example first (dashed purple line) and second (solid yellow line) electron trajectories for intensities of $4.0 \times 10^{13}$ Wcm$^{-2}$ (left column) and $1.2 \times 10^{14}$ Wcm$^{-2}$ (right column). The inset in (a) shows a zoom-in near time zero. (c) & (d) Simulated electron correlation maps. (e) & (f) Experimentally measured electron correlation maps.

**Figure 3** (a) The experimental $C_2H_2^{2+}/C_2H_2^+$ (circles) and $H^+/C_2H_2^+$ (squares) ratios as a function of the estimated peak laser intensity. (b) Normalised and vertically shifted $C_2H_2^{2+}$ ion momentum distributions as the intensity is increased from bottom to top.

## Acknowledgements

We acknowledge financial support from the Spanish Ministry of Economy and Competitiveness, through FIS2014-56774-R, FIS2014-51478-ERC, the "Severo Ochoa" Programme for Centres of Excellence in R&D (SEV-2015-0522), the Catalan Agencia de Gestió d'Ajuts Universitaris i de Recerca (AGAUR) with SGR 2014-2016, Fundació Cellex Barcelona, the European Union's Horizon 2020 research and innovation programme under grant agreement No 654148 Laserlab-Europe, the Marie Sklodowska-Curie grant agreements No. 641272 and 264666, COST Action MP1203 and COST Action XLIC. M.G.P. is supported by the ICFONEST programme, partially funded by COFUND (FP7-PEOPLE-2013-COFUND) and B.W. was supported by AGAUR with a PhD fellowship (FI-DGR 2013–2015). X.W. was supported by Chemical Sciences, Geosciences and Biosciences Division, Office of Basic Energy Sciences, Office of Science, U. S. Department of Energy under Grant No. DE-FG02-86ER13491. X.M.T. was supported by a Grand-in-Aid for Scientific Research (C24540421) from the Japan Society for the Promotion of Science and HA-PACS Project for advanced interdisciplinary computational sciences by exa-scale computing technology.

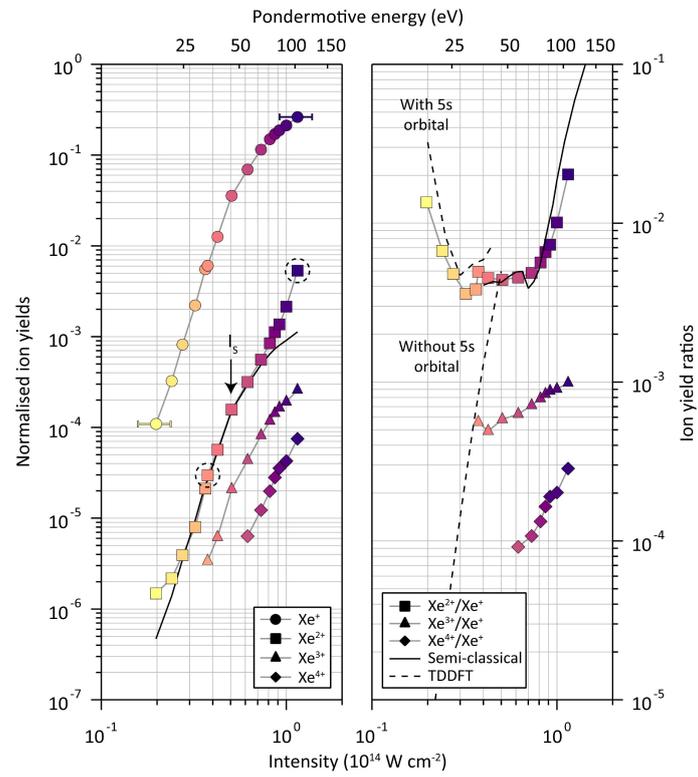

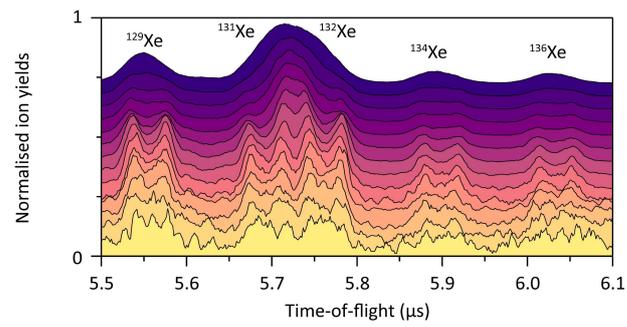

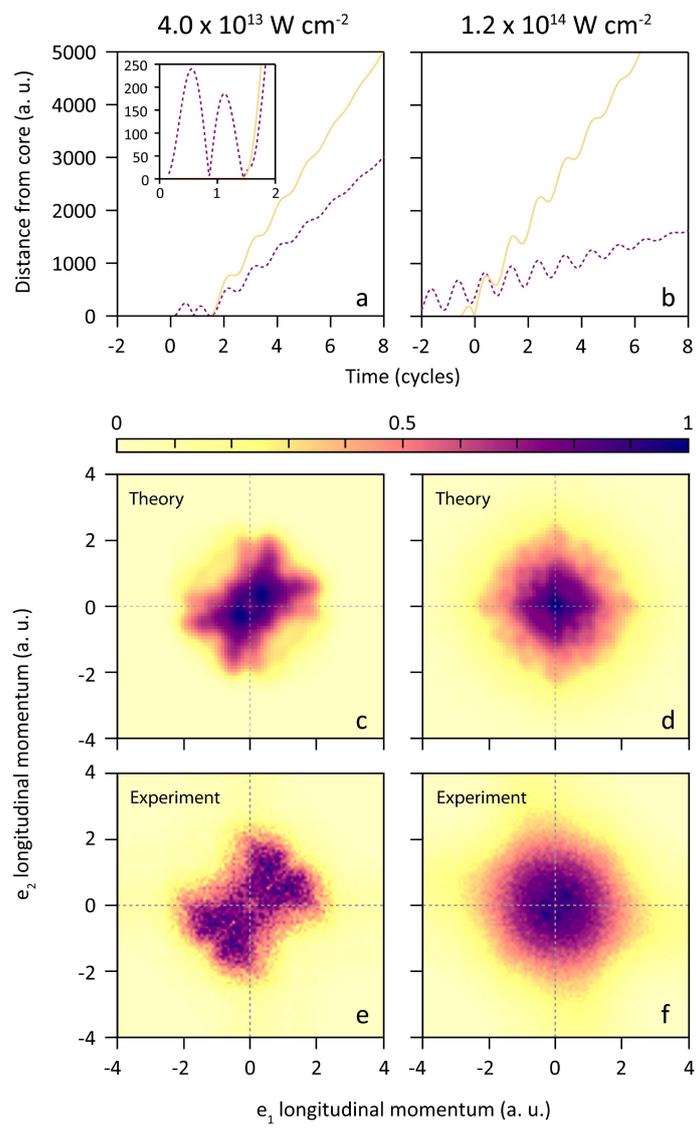

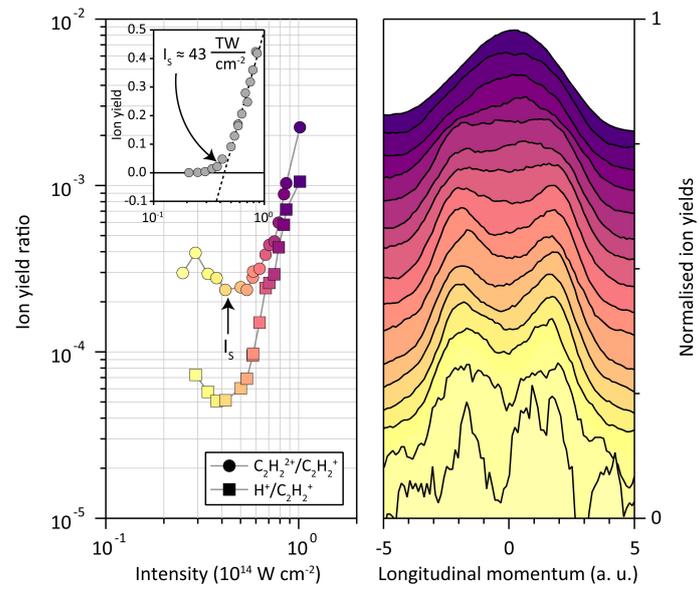